# Dynamic image potential at an Al(111) surface

I. D. White,[1] R. W. Godby,[2] M. M. Rieger[1] and R. J. Needs[1]

[1] *Theory of Condensed Matter Group, Cavendish Laboratory, Madingley Road, Cambridge CB3 0HE, United Kingdom*
[2] *Department of Physics, University of York, Heslington, York YO1 5DD, United Kingdom*

(November 7, 2017)

We evaluate the electronic self-energy $\Sigma(E)$ at an Al(111) surface using the *GW* space-time method. This self-energy automatically includes the image potential $V_{im}$ not present in any local-density approximation for exchange and correlation. We solve the energy-dependent quasiparticle equations to obtain surface state wavefunctions, and calculate the effective local potential experienced by electrons in the near-surface region. We find that $V_{im}$ for unoccupied states is due to correlation (not exchange). The image-plane position for interacting electrons is considerably closer to the surface than for the purely electrostatic effects felt by test charges, and, like its classical counterpart, is drawn inwards by the effects of atomic structure.

Electrons outside a metal surface experience a surface barrier which has the asymptotic form of an image potential $V_{im} = -1/4(z - z_o)$ [1], where $z_0$ is the effective edge of the metal. For interacting electrons this is an exchange-correlation (XC) effect, the quantum-mechanical analogue of the charge density redistribution which gives the classical image force. The form of the surface barrier is important for interpretation of low-energy electron diffraction [2] and scanning tunneling microscopy [3] experiments. In addition, surface and image states bound by $V_{im}$ can be directly observed by modern inverse photoemission and two-photon photoemission experiments [4].

The physics of the electron-surface interaction, and its transition from quantum-mechanical behaviour to the classical limit, has been the subject of a wealth of theoretical studies. It is well known that the effective potential of density-functional theory (DFT) within the local density approximation (LDA) (or gradient-corrected versions) fails to reproduce the image tail shape [5], which is the result of long-range many-body effects. The classical response of metal surfaces to an electric field or distinguishable test charges has been investigated using DFT [6–8], but the resulting image-plane position $z_c$ may differ from that experienced by electrons. The quantum-mechanical XC potential has been investigated by Eguiluz *et al.* for a jellium surface [9], while for real materials it has been shown that atomic structure beyond the jellium model plays an important role in determining both $z_c$ and the binding energy of image states [10] (whose existence requires the presence of a surface bandgap not exhibited by jellium).

In this Letter we present the results of a calculation of the non-local electronic self-energy $\Sigma(\mathbf{r}, \mathbf{r}', E)$ for an Al(111) surface, including its full variation with energy, evaluated within the *GW* approximation [11], which allows treatment of long-range correlation effects from first principles. This work is the first application of the *GW* space-time method, outlined in a recent Letter [12], to a system with a large supercell. In general surface states are poorly described in the LDA (and image states and resonances are entirely absent) and we therefore solve the quasiparticle (QP) equations self-consistently for one-electron-like excitation energies, without recourse to first-order perturbation theory. The effect of $\Sigma$ is interpreted in terms of the effective local potential $V_{loc}$ felt by QP states in a given energy range. $V_{loc}$ automatically contains the image potential and the form of the crossover from image to bulk behaviour which is crucial for states localised near the surface. We find $z_0$ (for quantum-mechanical electrons) at Al(111) to be significantly closer to the surface than $z_c$ (for external test charges). For Al(111) $z_0$ is shifted inwards relative to jellium by the atomic nature of the surface (as is seen for $z_c$). There has been ongoing controversy in the literature over the separate contributions of exchange and correlation [13,9,14], and we show that the image potential felt by excited-state electrons is not present in an exchange-only calculation.

In many-body theory, exchange and correlation are described by the self-energy $\Sigma(\mathbf{r}, \mathbf{r}', E)$, which is non-local and therefore in principle state-dependent, varies with energy, and is complex, containing information about the lifetimes of QP excitations. The non-local effects which give the image potential are present implicitly in the exact exchange-correlation potential $V_{xc}$ of Kohn-Sham DFT (as opposed to $V_{xc}^{LDA}$), but the surface barrier felt by excited states and the lifetimes of surface states are two features which cannot formally be addressed within DFT. The *GW* approximation for the self-energy, written in real space and time, is

$$\Sigma(\mathbf{r}, \mathbf{r}', t) = iG(\mathbf{r}, \mathbf{r}', t)W(\mathbf{r}, \mathbf{r}', t) \qquad (1)$$

where $G$ is the one-particle Green's function, and $W$ is the dynamically screened Coulomb interaction. This first-order diagram for $\Sigma$ has been shown to work well in a wide variety of studies of real materials, successfully predicting, for example, the band gaps of semiconductors and insulators [15] and the valence bandwidths of simple metals [16]. We follow the usual procedure of performing an LDA-DFT calculation to calculate a non-interacting approximation for $G$, and obtain $W$ within the random-phase approximation (RPA). We note that $W$ calculated at this level already contains image-like interactions, as RPA screening corresponds to time-dependent Hartree theory.

In the space-time method $\Sigma$ is constructed in real space and (imaginary) time, which is advantageous because the $GW$ self-energy is then a product rather than a convolution in reciprocal space and energy. In imaginary time, the structure of the many-body response functions is much smoother and thus well suited to numerical work. Fast Fourier transforms (FFTs) are used extensively to move between real and reciprocal space, and between imaginary time and energy. Our calculation begins with the formation of the Green's function $G(\mathbf{r},\mathbf{r}',i\tau)$ in real space and imaginary time. We then proceed via the non-interacting density response function and dielectric matrix to calculate $W$ without the use of any plasmon-pole approximation for frequency dependence. $\Sigma(\mathbf{r},\mathbf{r}',i\tau)$ is then formed and matrix elements in the LDA eigenfunction basis are computed, Fourier transformed to imaginary energy, and fitted to the multiple-pole form

$$\langle \psi_{n\mathbf{k}}|\Sigma(i\omega)|\psi_{n'\mathbf{k}}\rangle = a^0_{nn'\mathbf{k}} + \sum_{i=1}^{p} \frac{a^i_{nn'\mathbf{k}}}{i\omega - b^i_{nn'\mathbf{k}}}. \qquad (2)$$

This form facilitates analytic continuation to the real energy axis, and represents a highly controlled approximation since the accuracy of the fitting of the calculated self-energy can be directly monitored.

The system we study is an Al(111) surface. We employ a slab geometry with 5 layers of aluminium (sufficient to give, for example, a well-converged surface energy and $z_c$) and 8 layers of vacuum. The LDA calculation for the slab was undertaken with a $4\times4\times1$ k-point mesh and a high energy cutoff. In order to converge our surface barrier, the required parameters for the $GW$ calculation were found to be a plane-wave energy cutoff of 9 Ry (corresponding to a $5\times5\times60$ real space grid in the unit cell), and 243 imaginary time points with $\Delta t = \pi/10$ a.u. 300 bands (those up to an energy of 74 eV above $E_F$ in the LDA calculation) were included in $G$.

QP energies for bulk materials are usually evaluated within first-order perturbation theory, employing the assumption that the wavefunctions given by the solution of the QP equations,

$$\left(-\frac{1}{2}\nabla^2 + V_{ext}(\mathbf{r}) + V_H(\mathbf{r})\right)\Psi_{n\mathbf{k}}(\mathbf{r})$$
$$+ \int d\mathbf{r}' \Sigma(\mathbf{r},\mathbf{r}',E_{n\mathbf{k}})\Psi_{n\mathbf{k}}(\mathbf{r}') = E_{n\mathbf{k}}\Psi_{n\mathbf{k}}(\mathbf{r}), \qquad (3)$$

are sufficiently similar to their DFT Kohn-Sham (KS) counterparts, where $\Sigma$ is replaced by $V_{xc}$. The ability to solve the full QP equation for real materials is potentially important for highly inhomogeneous systems such as heterostructures or defects where the LDA may give qualitatively incorrect states.

In the case of the metal surface, the QP eigenfunctions will include surface and image states, bound by $V_{im}$, which will differ significantly from the LDA states. We therefore obtained QP eigenfunctions by diagonalising the QP Hamiltonian in the LDA eigenfunction basis at a trial energy, and then iterating the energy of the given QP state to self-consistency. The full energy-dependence of the self-energy matrix in the basis of LDA states is therefore required, making use of the space-time approach especially important, as a functional form for the energy dependence of $\Sigma$ is found. Even for the very inhomogeneous surface-slab system, two poles give a stable fit of extremely high quality (rms error 0.2%) for the diagonal matrix elements. Most off-diagonal matrix elements are zero by symmetry, with the (sparse) remainder similarly well described by the two-pole form. In Fig. 1 we show an unoccupied surface state below the vacuum level at $\Gamma$, obtained directly from solution of the QP equation. The weight in the near-surface region is significantly enhanced relative to the LDA state by the improved description of XC effects.

Although $\Sigma$ is non-local it can also be viewed as a state-dependent local potential. In particular, in the asymptotic limit far from the surface we expect the effect of $\Sigma$ to be that of an image potential, independent of QP energy [17]. Comparing the Hamiltonian for a local and non-local potential, it is clear that the state-dependent effective local XC potential $V_{loc}$ for a QP state is defined by

$$V_{loc}(\mathbf{r})\Psi_{QP}(\mathbf{r}) = \int d\mathbf{r}' \Sigma(\mathbf{r},\mathbf{r}',E_{QP})\Psi_{QP}(\mathbf{r}'), \qquad (4)$$

as used by Deisz *et al.* [18] in their study of a jellium surface. In the space-time method $\Sigma$ is obtained on the real energy axis in the form of matrix elements in the LDA eigenfunction basis. We therefore use the completeness relation to write

$$V_{loc}(\mathbf{r}) = \sum_n \langle \mathbf{r}|\psi_n\rangle\langle \psi_n|\Sigma(E_{QP})|\Psi_{QP}\rangle/\langle \mathbf{r}|\Psi_{QP}\rangle, \qquad (5)$$

where the sum is over all KS states $\psi_n$ at a given k-point. As the potential is not defined by Eq. 5 at the nodes of the QP state, we take a weighted average of the resulting $V_{loc}(\mathbf{r})$, according to $|\Psi_{QP}(\mathbf{r})|^2$, over a few states in a small energy range.

The resulting surface barrier, split into contributions from exchange and correlation, is shown in Fig. 2. This potential was calculated using four states at the $\Gamma$ point within 1 eV of the vacuum energy, and was well converged with the inclusion of 200 bands in the sum over off-diagonal matrix elements.

As expected, the bulk value of $V_{loc}$ is similar to $V_{xc}^{LDA}$, as the QP energy shift for states near $E_F$ in bulk

Al is rather small. Moving out through the surface, where $V_{xc}^{LDA}$ falls exponentially (as the density does), $V_{loc}$ crosses smoothly to the asymptotic image potential (shown with the best-fit $z_0$). This image potential has been modified to take account of the repeated slab geometry, which gives rise to two infinite series of image charges, but the resulting form becomes very similar to an isolated image potential within 10 a.u. of the surface. The form of the crossover is often treated in an *ad hoc* manner which is somewhat arbitrary for Al(111) as the classical $V_{im}$ and $V_{xc}^{LDA}$ do not meet. However it is interesting to note that the dynamic image potential limit which we have calculated in this work comes much closer to meeting $V_{xc}^{LDA}$.

Surface corrugation of the XC potential, which is well described by the LDA near the surface, quickly decays outside the surface. Our results are consistent with a surface-position-independent image plane, as argued and demonstrated for test charges by Finnis *et al.* [8]. The potential also proves to be almost entirely state-independent; in the case of the image tail this energy independence arises through cancelling effects from the spatial character of different states and the energy-dependence of $\Sigma$.

The 'classical' image-plane position $z_c$ has been calculated for jellium, and more recently for real materials, using a variety of techniques, generally based on the response of the ground-state charge density to external fields [6,7,19] and test charges [8]. The image-plane position has also been found through the self-consistent response of a modified effective potential [20]. These results, together with the $V_{xc}^{GW}$ value for dynamic electrons at a jellium surface obtained by Eguiluz *et al.* [9], are compared in Table I with $z_0$ for electrons at an Al(111) surface as calculated in this work, where the geometric edge is half a layer spacing outside the outermost crystal plane.

Our value for $z_0$ is closer to the surface than $z_c$ for Al(111) (as found for the jellium $z_0$ and $z_c$), and is closer than $z_0$ for jellium (as for the Al(111) and jellium $z_c$). The first trend can be thought of as resulting from the difference between the XC hole and the screening charge density caused by a point charge, and has been discussed for jellium by Eguiluz and Hanke [21] in terms of 'electron overlap effects'. The shift inwards for Al is however somewhat less than for jellium. The second trend is caused by the effects of atomic structure. Whereas the jellium model predicts that $z_0$ and $z_c$ should be a fixed distance from the geometric edge, $z_c$ is seen to be more closely tied to the position of the outermost layer, with the screening charge density centred just above the atoms. The value of $z_0$ for Al(111) suggests that the XC hole is similarly modified from the cylindrically symmetric case of jellium [22],(although the size of this shift is also somewhat smaller than that for $z_c$) and that $z_0$ will thus also tend to follow the position of the outermost crystal plane.

The issue of the physical origin of $V_{im}$ has provoked considerable controversy. Eguiluz *et al.* [9] found that the local potential at a jellium surface contained an exchange part with a $1/z^2$ dependence, and thus also ascribed the image tail to correlation, in contrast to earlier work by Harbola and Sahni [23], who obtained the image potential in $V_{xc}$ as the work done against a bare exchange hole. This discrepancy was explained more recently by Solomatin and Sahni [14] who showed analytically that the asymptotic exchange image form is given by a highly delocalised exchange hole in the semi-infinite metal, and predicted the observed faster decay of the exchange part in the case of a slab geometry. Thus, although technically exchange alone can provide the image tail for a genuinely semi-infinite metal, inclusion of exchange *and* correlation effects are clearly necessary to provide a correct physical (slab-width independent) description for electrons in the near-surface region.

In any case, $V_{xc}$ is felt only by electrons in occupied states, whereas our potential is that felt by excited states. The exchange part of our effective local potential shows exponential decay (as suggested by Almbladh and von Barth [13]), rather than power law behaviour. This is because the exchange part of the self-energy is given by $\Sigma_x = iGV$ where $G$ is evaluated at an infinitesimal positive time, and thus contains only occupied states. As a result (neglecting constant factors)

$$V_{loc}^{x}(\mathbf{r}) \propto \frac{\int d\mathbf{r}' \sum \psi_{occ}(\mathbf{r})\psi_{occ}^*(\mathbf{r}')\frac{1}{|\mathbf{r}-\mathbf{r}'|}\Psi_{QP}(\mathbf{r}')}{\Psi_{QP}(\mathbf{r})}. \quad (6)$$

Moving into the vacuum, the exponential decay of the occupied states in $G$ means that $V_{loc}^{x}$ must decay exponentially unless the denominator falls equally quickly. Inclusion of correlation is therefore essential to describe the image potential felt by unoccupied electronic states.

In conclusion, we have evaluated the electronic self-energy at an Al(111) surface within the $GW$ approximation. Calculation of the full energy-dependent self-energy and solution of the quasiparticle equations are made possible by use of the real-space imaginary-time method. The resulting effective local potential shows that correlation must be included to obtain the image potential for excited-state electrons such as those involved in photoemission or low-energy electron diffraction. The image-plane position for the many-electron system is closer to the surface than that for classical response to external fields or charges, and is also significantly modified by the atomic structure of the surface.

The authors acknowledge EPSRC support.

[1] We use Hartree atomic units ($\hbar=e=m_e=4\pi\epsilon_0=1$) throughout.
[2] J. Rundgren and G. Malmström, Phys. Rev. Lett. **38**, 836 (1977); P. J. Jennings and R. O. Jones, Phys. Rev. B **34**, 6695 (1986).
[3] G. Binnig *et al.*, Phys. Rev. B **30**, 4816 (1984); G. Binnig *et al.*, Phys. Rev. Lett. **55**, 991 (1985).
[4] F. Passek and M. Donath, Phys. Rev. Lett. **69**, 1101 (1992); U. Höfer *et al.*, Science **277**, 1480 (1997).
[5] N. D. Lang and W. Kohn, Phys. Rev. B **1**, 4555 (1970).
[6] N. D. Lang and W. Kohn, Phys. Rev. B **7**, 3541 (1973).
[7] J. E. Inglesfield, Surf. Sci. **188**, L701 (1987).
[8] M. W. Finnis *et al.*, J. Phys.: Condens. Matter **7**, 2001 (1995).
[9] A. G. Eguiluz, M. Heinrichsmeier, A. Fleszar, and W. Hanke, Phys. Rev. Lett. **68**, 1359 (1992).
[10] N. Garcia, B. Reihl, K. H. Frank, and A. R. Williams, Phys. Rev. Lett. **54**, 591 (1985).
[11] L. Hedin and S. Lundqvist, in *Solid State Physics 23*, edited by H. Ehrenreich, F. Seitz, and D. Turnbull (Academic, New York, 1969).
[12] H. N. Rojas, R. W. Godby, and R. J. Needs, Phys. Rev. Lett. **74**, 1827 (1995).
[13] C.-O. Almbladh and U. von Barth, Phys. Rev. B **31**, 3231 (1985).
[14] A. Solomatin and V. Sahni, Phys. Lett. A **212**, 263 (1996).
[15] M. S. Hybertsen and S. G. Louie, Phys. Rev. Lett. **55**, 1418 (1985); R. W. Godby, M. Schlüter, and L. J. Sham, Phys. Rev. Lett. **56**, 2415 (1986).
[16] J. E. Northrup, M. S. Hybertsen, and S. G. Louie, Phys. Rev. B **39**, 8198 (1989).
[17] J. C. Inkson, J. Phys. F: Metal Phys. **3**, 2143 (1973).
[18] J. J. Deisz, A. G. Eguiluz, and W. Hanke, Phys. Rev. Lett. **71**, 2793 (1993).
[19] S. C. Lam and R. J. Needs, J. Phys.: Condens. Matter **5**, 2101 (1993).
[20] P. A. Serena, J. M. Soler, and N. Garcia, Phys. Rev. B **37**, 8701 (1988).
[21] A. G. Eguiluz and W. Hanke, Phys. Rev. B **39**, 10433 (1989).
[22] J. E. Inglesfield and I. D. Moore, Solid State Comm. **26**, 867 (1978).
[23] M. K. Harbola and V. Sahni, Phys. Rev. B **39**, 10437 (1989).
[24] P. A. Serena, J. M. Soler, and N. Garcia, Phys. Rev. B **34**, 6767 (1986).

TABLE I. Image-plane positions for Al(111) surface in a.u., relative to the geometric edge, and for jellium with $r_s = 2.07$. Results for dynamic electrons (jellium calculations of Ref. [9], Al(111) this work) are compared with calculations based on response of the ground-state density by Lang and Kohn [6], Lam and Needs [19], Finnis [8], and Serena, Soler and Garcia [24,20] (see text).

|         | Dynamic    | LK   | LN   | Finnis | SSG  |
|---------|------------|------|------|--------|------|
| Jellium | 0.72±0.1   | 1.60 |      |        | 1.49 |
| Al(111) | 0.4±0.2    |      | 0.95 | 0.81   | 1.1  |

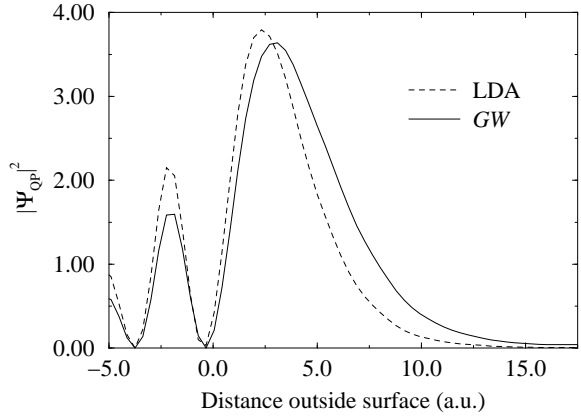

FIG. 1. Surface state QP wavefunction (full line), obtained by the iterative solution of the energy dependent QP equation at $\Gamma$, has weight transferred into the vacuum relative to the corresponding KS-LDA eigenfunction (dashed line).

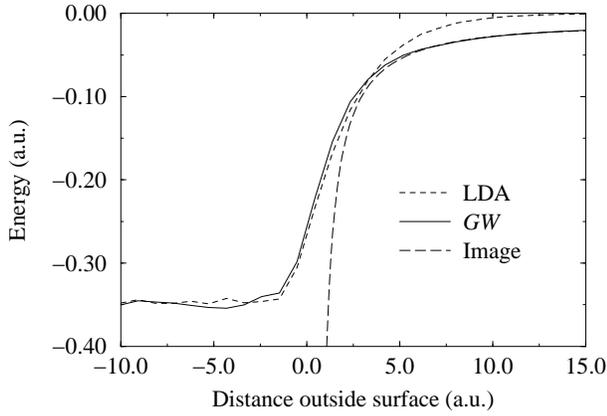

FIG. 2. Surface averaged effective local potential at Al(111) compared with $V_{xc}^{LDA}$. The XC potential calculated from $\Sigma$ crosses over to the classical image form in the vacuum (best-fit $V_{im}$ shown).

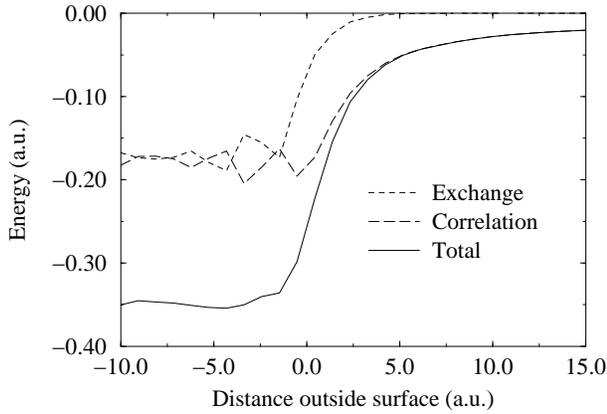

FIG. 3. Exchange and correlation contributions to the effective local potential for excited-state electrons at the Al(111) surface. The bare exchange part decays exponentially into the vacuum.